\documentstyle[12pt]{article}

\newcommand{\beq}[1]{\begin{equation} \label{#1} }
\newcommand{\eeq}    {\end{equation}}
\newcommand{\hhe}{\mbox{${\rm H}(^6{\rm He},\alpha)$}}
\newcommand{\kp}{\mbox{\ae}}
\newcommand{\arctg}{\mathop{\rm arctg}\nolimits}

\begin{document}
\pagenumbering{roman}

\mbox{}
\vspace{1cm}

\noindent{\large Russian Research Center "Kurchatov
Institute"}
\vspace{1cm}

\begin{tabbing}
\large A.L.Barabanov \` \large Preprint IAE-5811/2\\
\end{tabbing}
\vspace{2cm}

\noindent{\large DOES THE EXCITED STATE \\[\medskipamount]
OF THE $^3$H NUCLEUS EXIST\,?}
\vspace{1cm}

\noindent To be published in {\it Pis'ma ZETPh\/}
{\bf 61} n.1 (1995)
\vspace{5cm}

\centerline{\large Moscow -- 1994}

\newpage

\noindent {}
\vspace{0.5cm}

\noindent Key words: triton, neutron, deuteron, elastic
scattering, resonance, virtual state
\vspace{0.5cm}

\noindent The suggestion is made that the excited state
of the $^3$H nucleus
found out recently in the reaction \hhe\ ({\it Pis'ma JETPh\/}
{\bf 59} (1994) 301) has spin and parity~$1/2^+$ and the same
configuration that the ground one of~$^6$He. An amplitude of
electromagne\-tic transition to the triton ground state is
strongly suppressed, therefore the excited state cannot be
detected in radiative capture of neutrons by deuterons. It is
shown that in an elastic nd-scattering a resonance associated
with the exited state may be absent due to the destructive
interference of potential and resonant scattering phases.
\vspace{7cm}

\noindent \copyright\ Russian Research Center
"Kurchatov Institute", 1994

\newpage
\pagenumbering{arabic}

In Ref.\cite{1} the maximum in the cross section of the
reaction \hhe\ has been found and interpreted as the excited
state of the $^3$H nucleus with the energy
$E\,^*=7.0\pm0.3$~MeV
and the width $\Gamma\,^ * =0.6\pm0.3$~MeV.
This result is very
interesting as so far there was no evidence for triton
excited states~\cite{2}. There exists, however, the reasons
for skeptical attitude to the interpretation of~Ref.~\cite{1}.
The energy~$E\,^*$ is greater than the energy $E_b=6.26$~MeV
of the $^3$H nucleus break-up to neutron and deuteron
approximately by~0.7~MeV. But in the total cross section of
the
nd-interaction there is no maximum close to the energy~0.7~MeV
with the width of scale~$\Gamma\,^*$~\cite{3,4}. There are
also no anomalies in the radiative neutron-deuteron capture
near
the suggested excited level of triton~\cite{5}. A reference
made in Ref.~\cite{1} on the review~\cite{6} of theoretical
works
predicting a triton excited state lying over the break-up
threshold $n+d$ by~0.5~MeV is not quite correct as in
Ref.~\cite{6} the virtual state of nd-system with the energy
$\sim$~0.5~MeV with respect to the threshold $n+d$ is
really discussed. The purpose of this paper is to show that
the interpretation of Ref.~\cite{1} nevertheless can
be truthful.

We discuss at first a possible configuration of the excited
state of the $^3$H nucleus found out in~Ref.~\cite{1}. In the
framework of the naive shell model with the oscillator
potential
we have for the ground state~$0^+$ of the $^6$He nucleus
the following configuration. Two protons
and two neutrons are at 1s-state and form $\alpha$-particle,
while the remaining two neutrons are at 1p-state with zero
total angular momentum. Let
${\bf r}_1$ and ${\bf r}_2$ are position vectors of these neutrons
with respect to $\alpha$-particle, while a vector
${\bf r}_{12 }={\bf r}_1-{\bf r}_2$ determines the relative
position of the neutrons and
${\bf r}$ is a vector of their center mass with respect
to $\alpha$-particle. Then for the normalized oscillator
functions we have identity
\begin{eqnarray}
&\frac1{\sqrt{3}} \sum_m (-1)^m
\psi_{1pm}({\bf r}_1)\psi_{1p-m}({\bf r}_2)=\mbox{}&
\nonumber\\[\medskipamount]
&\mbox{}=\frac1{\sqrt{2}} \left(
\psi_{2s}({\bf r}) \psi_{1s}({\bf {\bf r}_{12}})-
\psi_{1s}({\bf r}) \psi_{2s}({\bf {\bf r}_{12}}) \right).&
\label{1}
\end{eqnarray}
The functions $\psi_{1pm}({\bf r}_i)$ in the left-hand part
are written
for the potential $\mu\omega^2r^2_i/2$, where
$\mu=\mbox{$m_nm_{\alpha}/(m_n+m_{\alpha})$}$ is a reduced mass of
neutron and $\alpha$-particle. In the right-hand part
one obtains the
functions $\psi_{ns}({\bf r}_{12})$ for the potential
$\mu_{12}\omega^2_{12}r^2_{12}/2$, where $\mu_{12}=m_n/2$ and
$\omega_{12}=\mbox{$\omega m_{\alpha}/(m_n+m_{\alpha})$}$, and
the functions $\psi_{ns}({\bf r})$ corresponding to the potential
$\mu_r\omega^2_rr^2/2$, where
$\mu_r=\mbox{$2m_nm_{\alpha}/(2m_n+m_{\alpha})$}$ and
$\omega_r=\mbox{$\omega(2m_n+m_{\alpha})/(m_n+m_{\alpha})$}$. So
according~Eq.(\ref{1}) the location of two neutrons
at 1p-level with
zero total angular momentum is equivalent to superposition of
1s- and 2s-states on the relative variables
${\bf r}_{12}$ and $ {\bf r}$. It was shown in~Ref.\cite{7} that
this simple picture agrees well with the real
configuration of the ground
state of the $^6$He nucleus that is a weakly bound system of three
particles \mbox{$\alpha+n+n$}.

Let us assume that the excited state of the \mbox{$p+n+n$} system
found out in~Ref.~\cite{1} has the same configuration~(\ref{1}) as
the ground one of the $^6$He nucleus (proton replaces
$\alpha$-particle). It will be obviously the excited state~$1/2^+$
of the $^3$H nucleus. In this model the ground state~$1/2^+$
corresponds to two neutrons at 1s-state with respect to a proton.
We have with that for the oscillator functions
\beq{2}
\psi_{1s}({\bf r}_1) \psi_{1s}({\bf r}_2)=
\psi_{1s}({\bf r}) \psi_{1s}({\bf {\bf r}_{12}}),
\eeq
with the same relationships between $\mu, \mu_{12}, \mu_r$ and
$\omega, \omega_{12}, \omega_r$ as above stated, replacing,
naturally, $m_{\alpha}$ by the proton mass $m_p$.

This hypothesis explains why the excited state does not manifest
in the reaction \mbox{${\rm d}({\rm n},\gamma)$}.
According to the quantum
numbers only M1-transition is possible but a leading term of
the transition amplitude equals zero due to the orthogonality
of 2s- and 1s-states. This is the same situation
as for \mbox{$2{\rm s}\to 1{\rm s}$} transition in a hydrogen atom,
where, by the way, a one-photon transition is less probable than
a two-photon one.

We turn now to the analysis of an elastic nd-scattering.
The total
spin of neutron and deuteron is 1/2 or 3/2, therefore one
distinguishes doublet and quartet channels in each partial
wave. The
scattering lengths in these channels are $a_2=0.65\pm0.03$~fm
and $a_4=6.34\pm0.02$~fm~\cite{2}.

In the absence of excited states of the $^3$H nucleus one should
expect that the neutron-deuteron scattering at the energies below
the deuteron break-up threshold \mbox{$E_d=2.23$ MeV} is purely
potential. A scattering length should thus have the scale of
potential radius. Let us assume that this is the case for
quartet channel. The radius of nd-potential can be very large due
to the deuteron diffuseness. A radial deuteron wave function
decreases
at large distances as \mbox{$\sim\exp(-\gamma r)$}, where
\mbox{$1/\gamma\sim 5$ fm}~\cite{9}. Adding the radius
of nucleon-nucleon interaction~(\mbox{$\sim 2$ fm}),
we take the value
$R=7$~fm for the radius of neutron-deuteron potential.
Choosing the
potential in the form of a spherical rectangular well and fitting
its depth in quartet channel to the value $a_4=6.34$~fm we obtain
$U_4=7.58$~MeV. There exists a bound 1s-state in such well
with the energy~-3.98~MeV, however, it can not be populated
in the quartet channel because of the Pauli principle.

Let us assume now that there is an excited state of triton of the
\mbox{$p+n+n$} type~(\ref{1}) in the doublet channel. This state
should be considered as a closed inelastic channel coupled with
an elastic one $n+d$. We shall trace how such inelastic
channel can
influence the observables in an elastic channel using the simplest
model of two-channel scattering~\cite{10,11}. In this model a
particle interacts with a two-state system; the energy of a
ground state is zero while the energy of an excited one
is~$\epsilon$. A
resonance holds when an incident particle passes into
a bound state and an excitation of internal system occurs.

We take all potentials in the form of spherical rectangular wells
with the radius $R=7$~fm. Let the depths of the wells in elastic
and inelastic channels are~$U^{(0)}_2$ and~$U^{(1)}_2$
correspondingly; the depth of the channel coupling potential
is~$W$.
The equations for the radial s-wave functions of elastic
$F^{(0)}(r)$
and inelastic $F^{(1)}(r)$ channels at $r<R$ are of the form
\beq{3}
\left\{\!\!
 \begin{array}{rcl}
d\,^2F^{(0)}/dr^2+(2mU^{(0)}_2/\hbar^2)F^{(0)}+
(2mW/\hbar^2)F^{(1)}+k^2F^{(0)}\!&\!\!=\!\!&\!0,\\
d\,^2F^{(1)}/dr^2+(2mU^{(0)}_2/\hbar^2)F^{(1)}+
(2mW/\hbar^2)F^{(0)}+k^2_1F^{(1)}\!&\!\!=\!\!&\!0,\\
 \end{array}
\right.
\eeq
where $k$ is a wave number in an elastic channel,
corresponding to an
energy $E=\hbar^2k^2/2m$ ($m$ is a reduced mass), and
$k_1=\mbox{$(2m(E-\epsilon)/\hbar^2)^{1/2}$}$ is a wave number in
an inelastic channel. If $E<\epsilon$, an inelastic channel
is closed, so $k_1=iq_1$, where
$q_1=\mbox{$(2m(\epsilon-E)/\hbar^2)^{1/2}$}$. Outside the
interaction region $r>R$ we have
\beq{4}
\left\{
 \begin{array}{rcl}
F^{(0)}(r)&=&\exp(i\delta_2(k)) \sin (kr+\delta_2(k))/k,\\
F^{(1)}(r)&=&-iS^{(1)}\exp(ik_1r)/2(kk_1)^{1/2},\\
 \end{array}
\right.
\eeq
where $\delta_2(k)$ is an elastic scattering phase in the doublet
channel.

The solutions of Eqs.(\ref{3}) in the region $r<R$, which are
regular in zero, are of the form
\beq{5}
\left\{
 \begin{array}{rcl}
F^{(0)}(r)&=&A\sin\kp r+A'\sin\kp 'r,\\
F^{(1)}(r)&=&A(\Delta/W)\sin\kp r-A'(W/\Delta)\sin\kp 'r,\\
 \end{array}
\right.
\eeq
where $\kp =\mbox{$(2m(U^{(0)}_2+E+\Delta)/\hbar^2)^{1/2}$}$,\\
\phantom{where} $\kp'=
\mbox{$(2m(U^{(1)}_2+E-\epsilon-\Delta)/\hbar^2)^{1/2}$}$,\\
$\Delta=
\mbox{$((U^{(0)}_2-U^{(1)}_2+\epsilon)^2/4+W^2)^{1/2}-
(U^{(0)}_2-U^{(1)}_2+\epsilon)/2$}$.
The matching of the functions~(\ref{4}) and~(\ref{5}) and their
derivatives at $r=R$ permits to find out the factors $A(E)$,
$A'(E)$, $S^{(1)}(E)$ and the elastic scattering
phase~$\delta_2(E)$.

A s-wave scattering phase can be
expressed through the logarithmic derivative
$\Phi_2(E)=\mbox{$R\,(dF^{(0)}/dr)/F^{(0)}$}$ of
an elastic channel function at the matching point
\beq{6}
\exp (2i\delta_2(E))=\exp (-2ikR)
\frac{\Phi_2(E)+ikR}{\Phi_2(E)-ikR}.
\eeq
If $E_2$ is the energy of a resonance in the doublet channel
then according to the usual definition~\cite{12}
\mbox{$\Phi_2(E_2)=0$}.
Writing the function $\Phi_2(E)$ near this energy in the form
$\Phi_2(E)=\mbox{$(E_2-E)/\gamma_2$}$, we come to the Breit-Wigner
description of a resonance with the reduced width~$\gamma_2$ and
the energy dependent width \mbox{$\Gamma_2(E)=2kR\gamma_2$}. If
the given parametrization of a logarithmic derivative holds for
the energies \mbox{$E\to 0$}, we have for the scattering length
$a_2=\mbox{$R(1-\gamma_2/E_2)$}$. We obtain a small value $a_2$ as
compared with~$R$ if \mbox{$\gamma_2\sim E_2$}. One follows from
the value \mbox{$E_2\simeq 0.7$ MeV}~\cite{1} that
\mbox{$\gamma_2\simeq 0.6$ MeV} and
\mbox{$\Gamma_2\simeq 1.3$ MeV}. We notice that the reduced width
of a purely potential resonance in the model of a spherical
rectangular well
$\gamma\,^{pot}=\hbar^2/mR^2\simeq 1.3$~MeV approximately twice
surpasses the found value of $\gamma_2$. Simultaneously the
width~$\Gamma_2$ is somewhat higher than the experimental
estimate~$\Gamma\,^*$~\cite{1}.

According with~Eq.(\ref{6}) an elastic scattering phase in
the doublet channel is the sum of a negative phase of potential
scattering $\delta\,^{pot}(E)=-kR$ and a positive phase of
resonant
scattering $\delta\,^{res}(E)=\mbox{$\arctg\,(kR/\Phi_2(E))$}$. If
the radius of nd-interaction is really so large as we have assumed
then at the energy \mbox{$E_2\simeq 0.7$ MeV} the potential phase
reaches the value \mbox{$\delta\,^{pot}(E_2)\simeq-1.05$}
comparable with the absolute value of the resonant phase
\mbox{$\delta\,^{res}(E_2)=\pi/2$}. As a result the total
phase~$\delta_2(E)$ does not pass through~$\pi/2$ in the
resonance. This is an explanation of the maximum absence in
the nd-scattering cross section near the energy~$E_2$.

We use the model above formulated for illustrative calculations.
Let $\epsilon$
equals the deuteron binding energy~2.23~MeV so the energetic
interval \mbox{$E<\epsilon$} where a purely elastic
scattering holds in the model coincides with the similar
interval in the nd-reaction.
An experimental value of scattering length in the doublet
channel can be easily reproduced using the remaining three free
parameters --- the depths of the potentials $U^{(0)}_2$,
$U^{(1)}_2$ and
$W$. Let, e.g., \mbox{$U^{(0)}_2=5$ MeV} and
\mbox{$U^{(1)}_2=U_4$}. Fitting $a_2$ to the value~0.65~fm we
found \mbox{$W=3.68$ MeV}. Then the logarithmic derivative
$\Phi_2(E)$ equals zero at \mbox{$E_2=0.76$ MeV}; the reduced and
total resonance widths are \mbox{$\gamma_2=0.57$ MeV} and
$\Gamma^0_2=\Gamma_2(E_2)=1.24$~MeV. The scattering cross sections
$\sigma_2(E)=\mbox{$(4\pi/3k^2)\sin^2\delta_2(E)$}$,
$\sigma_4(E)=\mbox{$(8\pi/3k^2)\sin^2\delta_4(E)$}$
in the doublet and
quartet channels and their sum --- the total s-wave cross section
are
shown in Fig.\ref{Fig.1}a. We see, that the doublet cross section
\begin{figure}
\vspace{10cm}
\caption{The s-wave cross section (a) and the squared
integral of the wave function
of inelastic channel (b) versus the energy $E$ calculated in the
model of nd-interaction; on fig.(a) the cross section
$\sigma_2(E)$ in doublet channel is shown by short strokes, the
cross section $\sigma_4 (E)$ in quartet channel --- by long
strokes, the total cross section --- by continuous line.
\label{Fig.1} }
\end{figure}
$\sigma_2(E)$ reaches the maximal values at the energies higher
than~1~MeV and slows down the drop of the total cross section in
agreement with the experiment~\cite{3,4}.  An increase of the
p- and d-wave contributions results in the same effect.

Thus, in the given model the resonance does not manifest as
a well expressed maximum
in an elastic channel because of the large
potential scattering phase. However, there is an enhancement of
the function $F^{(1)}(r)$ in an inelastic channel near
the energy $E_2$.
We notice that the probability of population of the triton excited
state in the reaction \hhe\ is proportional to the squared module
of matrix element of overlapping of the functions of
type~(\ref{1})
for the $^6$He and $^3$H nuclei. But the triton function of type~(\ref{1})
represents the function of the inelastic channel in the
reaction $n+d$. Thus, the enhancement of this function at the
energies near $E_2$ should correspond to the maximum in the cross
section of the reaction \hhe. For illustration we have calculated
the squared integral of the function $F^{(1)}$
\beq{7}
w(E)=\left(\int^R_0F^{(1)}(r)\,dr\right)^2
\eeq
in the our model. It is shown in Fig.\ref{Fig.1}b versus
the energy $E$.
Curiously, the position
and width of the maximum agree qualitatively with the
values of $E\,^*$ and $\Gamma\,^*$ found out in~Ref.\cite{1}.

We discuss, at last, an accordance between
the excited state~$1/2^+$ of the $^3$H nucleus and the virtual state
of \mbox{nd-system} in the doublet channel~\cite{6}. One says about
virtual
state if there is a pole in an elastic scattering
amplitude~(\ref{6}) in a bottom half-plane of complex values~$k$
on imaginary axis. If an energy of a resonance
$E_2=\mbox{$\hbar^2k^2_2/2m$}$ is small, the Breit-Wigner
approximation $\Phi_2(k)=\mbox{$\hbar^2(k^2_2-k^2)/2m\gamma_2$}$
may be true in some area of complex values~$k$ including zero.
Then the poles~$k_r$ of the amplitude~(\ref{6})
are the roots of square equation
$(k_r)_{1,2}=-ik_2(\Gamma^0_2/4E_2)\pm
k_2(1-(\Gamma^0_2/4E_2)^2)^{1/2}$.
We see, that in the given approximation virtual states correspond
to wide resonances $\Gamma^0_2>4E_2$ (both values~$k_r$ lie in
a bottom half-plane on imaginary axis). This is indeed the case in
a singlet channel of nucleon-nucleon scattering. The poles
corresponding to a resonance with a width~$\Gamma^0_2$ comparable
with~$E_2$, as considered here, lie on each side of
the imaginary axis.

In reality even at low energy~$E_2$ a deviation of~$\Phi_2(k)$
from the Breit-Wigner form can be built up rapidly when $k$ is
moving
away from the real axis. Therefore the pole on imaginary axis,
i.e., the virtual state, can correspond to the resonance with the
width \mbox{$\Gamma^0_2\sim E_2$}.
However, in the given model this is
not the case. Using the same parameters as were fixed above
we find by direct calculation the
poles of the amplitude~(\ref{6}) of doublet nd-scattering,
that is, zeros of the expression
\mbox{$\Phi_2(k)-ikR$}. Just
as the roots of square equation written out, they appear located
symmetrically to the imaginary axis in a bottom half-plane. We
have for the complex energies
$(E_r)_{1,2}=\mbox{$\hbar^2(k_r)^2_{1,2}/2m$}$ on a non-physical
sheet ${\rm Re}(E_r)_{1,2}=\mbox{$0.54$ MeV}$,
${\rm Im}(E_r)_{1,2}=\mbox{$\pm 0.76$ MeV}$. Thus, in the given
model the resonance in nd-scattering (see the Fig.\ref{Fig.1}b) is
correlated with the poles of S-matrix near the imaginary axis.
The pole energies on a non-physical sheet are of
scale~$\sim$~0.5~MeV. Taking into account an extreme simplicity of
the model under consideration one may say about qualitative
agreement with the results of the works discussed
in~Ref.~\cite{6}.

So, in this paper it was shown that the interpretation of
the maximum
in the reaction \hhe\ as the excited state of the nucleus $^3$H
does not contradict to the available
experimental data and theoretical conceptions about the poles of
S-matrix in nd-interaction.
The excited state does not apparently manifest
in the elastic nd-scattering because of the destructive
interference of potential and resonant
scattering phases. The reason for this effect is
the anomalously large radius of nd-potential, related with
the deuteron diffuseness.
According to the advanced assumption about the structure of the
excited state the cross section of the reaction
\mbox{d(n,$\gamma$)} is suppressed because of a strong prohibition
of M1-transition to the triton ground state. While
the hypothesis about similarity of the structure of the
excited state of the $^3$H nucleus to that of the ground one of
the $^6$He nucleus naturally explains the sensitivity of the cross
section of the reaction \hhe\ to this state found out
in~Ref.~\cite{1}.
{\samepage

}

The main conclusion is one should follow the energy dependence of
the wave functions of the inelastic channels that are close on
structure to the configuration~(\ref{1}) in the consistent
three-body calculation of nd-scattering. An estimate of the
possibility of
triton excitation in an inelastic electron scattering due to
E0-transition and an expansion of the analysis on the system p$+$d
and $^3$He nucleus are of the obvious interest.

\vspace{\baselineskip}

The author is grateful to D.V.Alexandrov, E.Yu.Nikolskii,
B.G.Novatskii, D.N.Stepanov, B.V.Danilin, M.V.Zhukov and
D.V.Fedorov for useful discussions. The work was supported
in part by International Scientific Foundation.

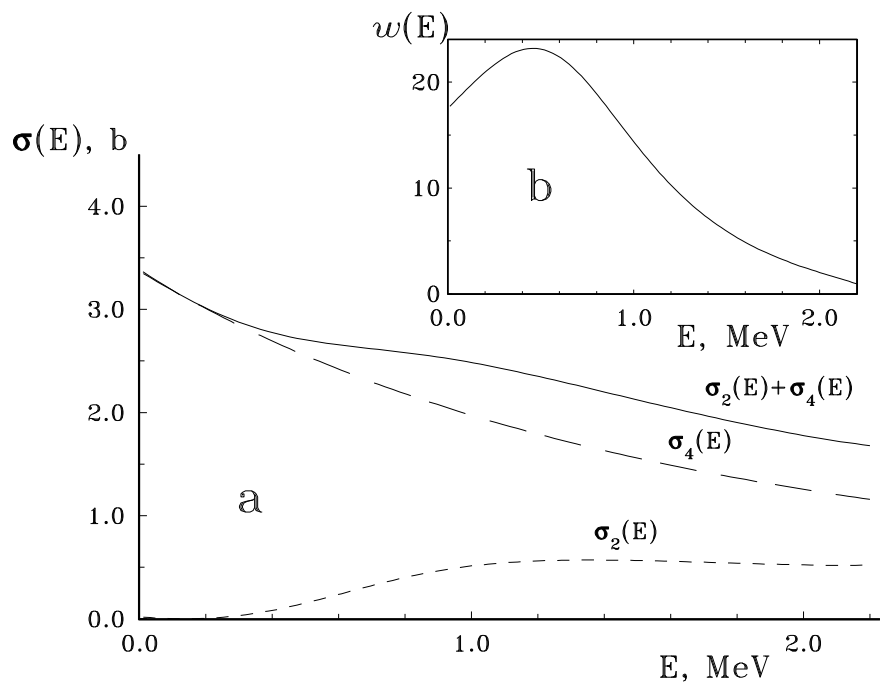
\end{document}